\documentclass[
 reprint,
superscriptaddress,
 amsmath,amssymb,
prl,
floatfix,
]{revtex4-1}
\usepackage{dcolumn}
\usepackage{bm}
\usepackage{geometry}
\usepackage{graphicx} 

\usepackage{float}
\usepackage{multirow}
\usepackage{booktabs}
\usepackage{array} 
\usepackage{paralist} 
\usepackage{verbatim} 
\usepackage{rotating}

\usepackage{fancyhdr} 
\usepackage{threeparttablex}
\usepackage{amsmath, amsthm}
\usepackage{afterpage}
\usepackage[nottoc,notlof,notlot]{tocbibind} 

\input{Qcircuit.tex}  

\begin{document}

\title {Quantum circuits for solving linear systems of equations}
\author {Yudong~Cao}\affiliation{ Department of Mechanical Engineering, Purdue University} 
\author{Anmer~Daskin}\affiliation{Department of Computer Science, Purdue University} 
\author{Steven~Frankel}\affiliation{ Department of Mechanical Engineering, Purdue University} 
\author{Sabre~Kais}\email{Corresponding author:kais@purdue.edu}\affiliation{Department of Chemistry, Physics and Birck Nanotechnology Center,Purdue University, West Lafayette, IN 47907 USA}

\begin{abstract}

Recently, it is shown that quantum computers can be used for obtaining certain information about the solution of a linear system $A\vec{x}=\vec{b}$ exponentially faster than what is possible with classical computation. 
Here we first review some key aspects of the algorithm from the standpoint of finding its efficient quantum circuit implementation using only elementary quantum operations, which is important for determining the potential usefulness of the algorithm in practical settings. Then we present a small-scale quantum circuit that solves a $2\times 2$ linear system. The quantum circuit uses only 4 qubits, implying a tempting possibility for experimental realization. Furthermore, the circuit is numerically simulated and its performance under different circuit parameter settings is demonstrated.
\end{abstract}
\maketitle

Quantum computers exploit quantum mechanical phenomena such as superposition and entanglement to perform computations. Because they
 compute in ways that classical computers cannot, for certain problems such as factoring large numbers \cite{shor94} and simulation of quantum systems \cite{lloyd96, abrams99, abrams97,traub10,zhang10,kais08,alan05,lidar99,nori11a,nori11b,nori05,nori09,dowling06,veis10,veis11} quantum algorithms provide exponential speedups over their classical counterparts. Recently, Harrow, Hassidim and Lloyd \cite{lloyd09} proposed a quantum algorithm for obtaining certain information about the solution $\vec{x}$ of a linear system $A\vec{x}=\vec{b}$. We will proceed by first giving a summary of the algorithm, followed by some remarks on the key aspects of the algorithm related to its efficient quantum circuit implementation. Then we present an example quantum circuit in order to encourage the experimental effort on implementing the algorithm.
$\quad$\\
$\quad$\\
\noindent\centerline{\it The algorithm}

Suppose an operator $A$ can be represented as an $N\times N$ Hermitian matrix with a spectral decomposition of $A=\sum_j\lambda_j|u_j\rangle\langle{u_j}|$ (the non-Hermitian cases can be accounted for by some simple modifications of the algorithm, see \cite[Sec.\ 4, Appendix A]{lloyd09}). The condition number is defined as $\kappa=\max_j|\lambda_j|/\min_j|\lambda_j|$. Without loss of generality we assume $\kappa^{-1}\le\lambda_j\le 1$ for all $j$. 

\begin{figure*}
\centerline{
\Qcircuit @C=0.6em @R=0.6em @!R{
\lstick{Anc.} & & |0\rangle\qquad & \qw & \qw & \qw & \qw & \qw & \qw & \qw & \qw & \qw & \qw & \qw & \qw & \qw & \qw & \gate{R_y(\tilde\theta_j)} & \qw & \qw & & |\alpha_j\rangle \\
\lstick{Reg.L} & & |0\rangle\qquad & {/} \qw & \qw & \qw & \qw & \qw & \qw & \qw & \qw & \multigate{2}{U_\lambda} & \qw & & & |\tilde\theta_j\rangle\quad & & \ctrl{-1} & \multigate{4}{U^\dagger} & \qw & & |0\rangle \\
\lstick{Anc.} & & |0\rangle\qquad & {/} \qw & \qw & \qw & \qw & \qw & \qw & \qw & \qw & \ghost{U_\lambda} & {/} \qw & \qw & \qw & \qw & \qw & \qw & \ghost{U^\dagger} & \qw & & |0\rangle \\
\lstick{Reg.C} & & |0\rangle\qquad & {/} \qw & \gate{W} & \ctrl{1} & \gate{FT^\dagger} & \qw & & \makebox[0.15cm]{}|\lambda_j\rangle\quad & & \ghost{U_\lambda} & {/} \qw & \qw & \qw & \qw & \qw & \qw & \ghost{U^\dagger} & \qw & & |0\rangle \\
\lstick{Reg.B} & & |b\rangle\qquad & {/} \qw & \qw & \multigate{1}{{e^{iAt}}} & \qw & \makebox[0.5cm]{}\sum_j\beta_j|u_j\rangle & & & & & \qw & \qw & \qw & \qw & \qw & \qw & \ghost{U^\dagger} & \qw & & \makebox[1.8cm]{}|b\rangle=\sum_{j}\beta_j|u_j\rangle \\
\lstick{Anc.} & & |0\rangle\qquad & {/} \qw & \qw & \ghost{e^{iAt}} & \qw & {/} \qw & \qw & \qw & \qw & \qw & \qw & \qw & \qw & \qw & \qw & \qw & \ghost{U^\dagger} & \qw & & |0\rangle
}}
\caption{Overview of the quantum circuit for solving the linear system $A\vec{x}=\vec{b}$. Each label $Anc.$ represents an ancilla register. $Reg.$ labels each register that stores (intermediate or final) computation results. $W$ is the Walsh-Hadamard transform which applies Hadamard transform on every qubit of the register. $FT$ represents quantum Fourier transform. The circuit for $FT$ is well known~\cite{nielsen00}. $U_\lambda$ is the subroutine that computes the state $|\tilde\theta_j\rangle$ with $\tilde\theta_j$ approximating $\theta_j=2\text{arcsin}(C/\lambda_j)$ for the eigenvalues of $A$ encoded in the states $|\lambda_j\rangle$. $U^\dagger$ represents the inverse of all the operations before the controlled $R_y$ rotation. For small rotation angles in $R_y$, the final state of the top ancilla bit is $|\alpha_j\rangle$, which approximates $\sqrt{1-{C^2}/{\lambda_j^2}}|0\rangle+{C}/{\lambda_j}|1\rangle$ up to $\varepsilon$.}
\label{fig:general_circuit}
\end{figure*}
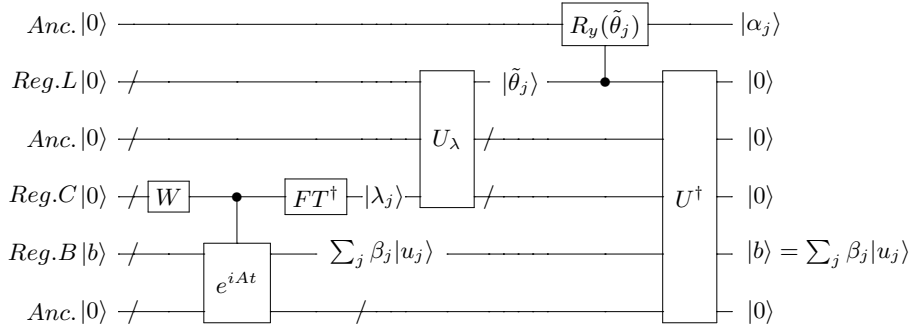

The general quantum circuit for the algorithm is shown in Fig.\ \ref{fig:general_circuit}. The right hand vector is encoded in the quantum state $\vec{b}$, which has an expansion $|b\rangle=\sum_j\beta_j|u_j\rangle$ in the eigenbasis of $A$. The algorithm starts by a well-known phase estimation subroutine, which involves applying the controlled unitary $U=e^{iAt}$ on $|b\rangle$ for a superposition of different $t$ values. After phase estimation we obtain a state that is approximately $\sum_j\beta_j|\lambda_j\rangle|u_j\rangle$ (Fig.\ \ref{fig:general_circuit}). Here $|\lambda_j\rangle$ is a state that encodes an approximation to the eigenvalue $\lambda_j$ \cite[Ch. 5]{nielsen00}.

The next stage of the algorithm is intended to bring the state of the system to be proportional to $\sum_j\beta_j\lambda_j^{-1}|u_j\rangle\otimes|\text{Anc.}\rangle$. Here $|\text{Anc.}\rangle$ is some state of ancilla qubits. The ancillas decouple from that of the subset of qubits in the state $\sum_j\beta_j\lambda_j^{-1}|u_j\rangle$, which is proportional to the solution $|x\rangle\propto A^{-1}|b\rangle$. To achieve this transformation, introduce an ancilla qubit initialized at $|0\rangle$ and use the $|\lambda_j\rangle$ states after phase estimation (Fig.\ \ref{fig:general_circuit}) to perform a controlled $Y$-rotation $R_y(\theta_j)=e^{-i\theta_j Y/2}$ ($Y$ is the Pauli Y operator) onto the ancilla qubit such that the state of the system is brought to 
\begin{equation}\label{eq:2}
\sum_j\left(\sqrt{1-\frac{C^2}{\lambda_j^2}}|0\rangle+\frac{C}{\lambda_j}|1\rangle\right)\beta_j|\lambda_j\rangle|u_j\rangle
\end{equation}
\noindent{}with the rotation angles $\theta_j=2\text{arcsin}(C/\lambda_j)$. Here the constant $C\le\min_j|\lambda_j|=O(1/\kappa)$. 

The final step of the algorithm is to apply the inverse of the phase estimation subroutine at the beginning and transform the register $|\lambda_j\rangle$ back to $|0\rangle$, thus transforming the system to $\sum_j\big(\sqrt{1-C^2/\lambda_j^2}|0\rangle+(C/\lambda_j)|1\rangle\big)\beta_j|0\rangle|u_j\rangle$. A projective measurement on the first ancilla qubit, when measured to be $|1\rangle$, will collapse the final state of $Reg.B$ (Fig.\ \ref{fig:general_circuit}) to the desired state
\begin{equation}
\sum_jC\frac{\beta_j}{\lambda_j}|u_j\rangle\propto|x\rangle
\end{equation}
\noindent{}with probability of $\sum_j|\beta_j|^2\cdot|C/\lambda_j|^2$, which scales as $O(1/\kappa^2)$.
$\quad$\\
$\quad$\\
\noindent\centerline{\it Discussion}
$\quad$\\
Here we review some key aspects of the algorithm related to finding its efficient quantum circuit implementation using only elementary operations.

A detailed complexity analysis in \cite{lloyd09} shows that the algorithm runs in $O(\log(N)\kappa^2/\epsilon)$ time where $\epsilon$ is the total error in the output state $|x\rangle$. The complexity, or the cost scaling of the quantum algorithm in terms of $\kappa$ and $\epsilon$ are proven to be optimal \cite[Sec.\ 5, Appendix A]{lloyd09}, while in cases such as $A$ being symmetric positive definite, the best classical algorithm Conjugate Gradient has a scaling $O(N\sqrt{\kappa}\log(1/\epsilon))$, see Ref.\ \cite{shewchuk94}. Hence the most useful application of the algorithm is limited to situations where neither $\kappa$ nor $1/\epsilon$ is large \cite{childs09}.

A major strength of the algorithm is that under certain conditions (which we will discuss next) it finds the solution $|x\rangle$ with $O(\log N)$ cost, while any classical algorithm requires at least $O(N)$ effort to write down the answer $\vec{x}$. Because the solution is encoded in the quantum state $|x\rangle=\sum_{i=1}^Nx_i|i\rangle$ and obtaining all the values of $x_i$ still requires $O(N)$ effort, the application of the algorithm is further limited to cases where we are only interested in certain feature of the solution that is represented by an expectation value $\langle{x}|M|x\rangle$ for some operator $M$.

With the above constraints, in order to retain the $O(\log N)$ cost scaling the algorithm assumes that the following subroutines are all efficient (with cost scaling $O(\text{poly}(\log N,\kappa,1/\epsilon))$):
$\quad$\\
$\quad$\\
\noindent{\bf State preparation of $|b\rangle$.} In general it is known that to prepare an arbitrary quantum state in an $N$-dimensional Hilbert space one needs $O(\text{poly}(N))$ elementary gates \cite{nielsen00}. This is immediately related to decomposing an arbitrary unitary operation to a product of elementary quantum gates (such as {\sf CNOT} gates and single-qubit rotations) since preparing any state $|b\rangle$ requires a unitary $U$ such that (for example) the transformation $U|0\cdots 0\rangle=|b\rangle$ serves as a preparation of $|b\rangle$ from $|0\cdots 0\rangle$, a state that is easy to prepare. There have been improved schemes for unitary gate decompositions \cite{MVBS04,BVMS05,PB11}, however there is no general scheme that will breach the $\text{poly}(N)$ bound in the cost scaling. Regardless of this fundamental limitation, there are particular types of states that are shown to be efficiently preparable.

One notable example is the case considered independently by Zalka \cite{Z98}, Grover and Rudolph \cite{GR02}, Kaye and Mosca \cite{KM04} where the state $|b\rangle=\sum_ib(i)|i\rangle$ corresponding to an efficiently integrable function $b(x)$ can be efficiently prepared. The initial motivation in \cite{Z98} is to encode a continuous basis wavefunction into a quantum state in preparation for quantum simulation. This idea is then extended in \cite{WKA09} for generating a wider variety of single- and multi-particle eigenstates for quantum simulation.

While the state generation schemes in \cite{Z98,GR02,KM04,WKA09} are cast in the standard gate model of quantum computation, Aharonov and Ta-Shma \cite{AT03,AT07} consider quantum state generation by simulating Hamiltonians of adiabatic evolutions that correspond to slowly varying Markov chains. According to \cite[Sec.\ 4]{AT03}, for a Markov chain described with a matrix $M$ acting on probability distributions over the state space $\Omega$, for a limiting distribution $\pi=\lim_{t\rightarrow\infty}M^tp$ with $p$ being the initial distribution, if $M$ is row computable and for any $i,j\in\Omega$, $M_{ij}\pi_i=M_{ji}\pi_j$ and $\pi_i/\pi_j$ can be efficiently approximated, then the Hamiltonian corresponding to $M$, defined as $H_M=I-\Lambda M\Lambda^{-1}$ where $\Lambda$ is a diagonal operator with $\sqrt{\pi_i}$ at the $i^\text{th}$ diagonal position, has its ground state being $\sum_i\sqrt{\pi_i}|i\rangle$. Because one can efficiently simulate an adiabatic evolution starting at a Hamiltonian corresponding to a simple Markov chain and ending at a Hamiltonian corresponding to $H_M$ (See \cite[Lemma 2 and 3]{AT03}), the state $\sum_i\sqrt{\pi_i}|i\rangle$ can thus be efficiently prepared. 
$\quad$\\
$\quad$\\
\noindent{\bf Hamiltonian simulation $e^{-iAt}$.} The problem of Hamiltonian simulation has been extensively studied. For a general non-sparse $N\times N$ Hamiltonian $H$, it is shown in \cite{childs10} that it is not possible to simulate $e^{-iHt}$ in $\text{poly}(\|Ht\|,\log N)$ time, which sets a fundamental limitation for the currently known Hamiltonian simulation schemes. In general, however, it is of greater interest to simulate sparse Hamiltonians efficiently. Particularly if $H$ is 1-sparse ($s$-sparse means that every row and column of the Hamiltonian has at most $s$ non-zero entries), $e^{-iHt}$ can be implemented with $O(1)$ elementary operations \cite{childs03,AT03}. In general, an $s$-sparse Hamiltonian can be decomposed as a sum of $O(s^2)$ 1-sparse Hamiltonians efficiently \cite{AT03,berry07}. Since the initial work by Lloyd \cite{lloyd96} for the case of time-independent local Hamiltonians, there has been several simulation algorithms formulated using product formulas \cite{berry07,zhang10,wiebe11} and improved using linear combinations of unitary operators \cite{wiebe12}, yielding a cost scaling that is $\text{poly}(\log N, 1/\epsilon)$ and almost linear in $\|Ht\|$. 

Alternatively, simulation algorithms using quantum walks \cite{childs09b,berry12} have cost scaling $O(\|Ht\|/\sqrt{\epsilon})$, which is strictly linear in $\|Ht\|$. None of the algorithms presented so far is able to show $O(\log(1/\epsilon))$ scaling in $1/\epsilon$ except for special cases where $H$ has a specific structure such as being proportional to a discrete Laplacian in any finite dimension \cite{cao13}. 

However, building on prior works \cite{berry07,cleve09,berry12} recently it has been shown in \cite{berry13} that to simulate $e^{-iHt}$ for an $s$-sparse Hamiltonian requires only $O(s^2\|Ht\|\text{poly}(\log N,\log(1/\epsilon)))$, breaching the limitation of previous algorithms on the scaling in terms of $1/\epsilon$. The issue of Hamiltonian simulation in $O(\log(1/\epsilon))$ is important because classical simulations of quantum systems. while suffering from the exponential scaling in $n=\log N$. enjoy $O(\log(1/\epsilon))$ scaling. The possibility of achieving the same $O(\log(1/\epsilon))$ scaling in the quantum regime could help answering open questions in numerical analysis \cite{wiebe12}.
$\quad$\\
$\quad$\\
\noindent{\bf Eigenvalue inversion $U_\lambda$.} To accurately transform the state of the system to \eqref{eq:2} requires a non-unitary operation. It is shown in \cite{cao13} that using quantum circuits to simulate classical root finding subroutines, the state $|\theta_j\rangle$ approximating $\theta_j$ up to error $\varepsilon$ can be prepared with $O(\text{polylog}(1/\varepsilon))$ cost.

$\quad$\\
\centerline{\it Example: solving a $2\times 2$ system}
$\quad$\\
Here we present a 4-qubit quantum circuit that solves the smallest meaningful instance of the problem: a $2\times 2$ system. The purpose for this example is for illustrating the algorithm and for potential experimental implementation using currently available resource. Hence simplifications with respect to the general quantum circuit implementation discussed in the previous section are in order. In the example $|b\rangle$ is a one-qubit state, which is easy to prepare. The quantum circuit for realizing Hamiltonian simulation $e^{-iAt}$ with elementary operations is found via a heuristic algorithm developed in some previous works \cite{daskin11a,daskin11b}. There is no performance guarantee for the efficiency of the heuristics for large matrices but for our purpose in this example they are sufficient. In order to simplify the eigenvalue inversion subroutine $U_\lambda$ (Fig.\ \ref{fig:general_circuit}), $A$ is chosen such that it has eigenvalues that are powers of $2$, so that the phase estimation subroutine will generate states that exactly encode the eigenvalues, making it simple to find their reciprocals. 

As we have shown previously in the general case, the mapping from $\sum_j\beta_j|\lambda_j\rangle|u_j\rangle$ to $\sum_j\beta_j\lambda_j^{-1}|\lambda_j\rangle|u_j\rangle\propto|x\rangle$ should in principle use controlled $Y$ rotation with angle $\theta_j=2\text{arcsin}(C/\lambda_j)$. Here we assign $C$ such that the small-angle approximation $\text{arcsin}(C/\lambda_j)\approx C/\lambda_j$ could reasonably hold. Although the inversion scheme introduced in the example is purely \emph{ad hoc}, with additional qubits in the $Reg.C$ (Fig,\ \ref{fig:general_circuit}) the implementation of $U_\lambda$ as described in \cite{cao13} is possible. 

\begin{figure*}
\[
\Qcircuit @C=.8em @R=0.2em @!R 
{
\lstick{|x_1\rangle} & \qw & \qw & \qw & \qw & \qw & \qw & \qw & \qw & \qw & \qw & \qw & \gate{R_y(\frac{2\pi}{2^r})} & \gate{R_y(\frac{\pi}{2^r})} & \qw & \qw & \meter & \makebox[0.5cm]{}|m\rangle \\
\lstick{|x_2\rangle} & \gate{H} & \qw & \ctrl{2} & \qw & \qswap & \qw & \ctrl{1} & \gate{H} & \qw & \qw & \qswap & \ctrl{-1} & \qw & \multigate{2}{U^\dagger} & \qw \\
\lstick{|x_3\rangle} & \gate{H} & \ctrl{1} & \qw & \qw & \qswap \qwx & \gate{H} & \gate{S^\dagger} & \qw & \qw & \qw & \qswap \qwx & \qw & \ctrl{-2} & \ghost{U^\dagger} & \qw \\
\lstick{|x_4\rangle} & \qw & \gate{\text{exp}(iA\frac{t_0}{4})} & \gate{\text{exp}(iA\frac{t_0}{2})} & \qw & \qw & \qw & \qw & \qw & \qw & \qw & \qw & \qw & \qw & \ghost{U^\dagger} & \qw \gategroup{4}{2}{4}{9}{2.5em}{_\}} \\
& & \makebox[3.6cm]{}\raisebox{-2.2ex}{\text{\small Phase estimation}} &  & & & & \\
&
}
\]
\caption{The example quantum circuit for solving the $2\times{2}$ linear system $A\vec{x}=\vec{b}$. Here $|x_1\rangle$, $|x_2x_3\rangle$ and $|x_4\rangle$ correspond respectively to the top ancilla qubit, register $C$ and register $B$ in Fig.\ \ref{fig:general_circuit}. We let $t_0=2\pi$ and $r$ is parameter that ranges between $\log_2(2\pi)$ and $\log_2(\pi/\omega)$ with $\omega$ being the minimum angle that can be resolved. Here $U^\dagger$ represents the inverse of all the operations before the $R_y$ rotations. $r>0$ is a parameter that will determine the final state probability. Initially $|x_4\rangle=|b\rangle$ and $|x_1\rangle$, $|x_2\rangle$ and $|x_3\rangle$ are all $|0\rangle$.}
\label{fig:circuit_2by2}
\end{figure*}

We let \begin{equation}\label{eq:Axb}
A=\frac{1}{2}\begin{pmatrix}
3 & 1 \\
1 & 3 \\
\end{pmatrix};
\vec{b}=\begin{pmatrix}
{b_1} \\ {b_2} \\
\end{pmatrix}.\end{equation}The circuit for solving the linear system is shown in Fig.\ \ref{fig:circuit_2by2}. Assuming $|b_1|^2+|b_2|^2=1$, the vector $\vec{b}$ can be encoded in the state $|b\rangle=b_1|0\rangle+b_2|1\rangle$. The eigenvalues of $A$ are $\lambda_1=1$ and $\lambda_2=2$ with corresponding eigenvectors $|u_1\rangle$ and $|u_2\rangle$. Note that $\lambda_1$ and $\lambda_2$ can be accurately encoded by $|x_2x_3\rangle=|01\rangle$ and $|x_2x_3\rangle=|10\rangle$ respectively. Therefore after the phase estimation the state of the 3-qubit system $|x_2x_3x_4\rangle$ reads $\beta_1|01\rangle|u_1\rangle+\beta_2|10\rangle|u_2\rangle$ where $\beta_1$ and $\beta_2$ are expansion coefficients of $|b\rangle$ in $A$'s eigenbasis. 

To obtain the state $|\theta_j\rangle$ for the eigenvalue inversion (Fig.\ \ref{fig:general_circuit}) we use an \emph{ad hoc} method that does not require any ancilla qubits. We first apply a {\sf SWAP} gate between $|x_2\rangle$ and $|x_3\rangle$ so that the 3-qubit system $|x_2x_3x_4\rangle$ is transformed to the state $\beta_1|10\rangle|u_1\rangle+\beta_2|01\rangle|u_2\rangle$. We can now interpret $|x_2x_3\rangle=|10\rangle$ as the state encoding the inverted eigenvalue $2\lambda_1^{-1}=2$ and $|x_2x_3\rangle=|01\rangle$ as that encoding $2\lambda_2^{-1}=1$. In other words, after the {\sf SWAP} gate following the phase estimation (Fig.\ \ref{fig:circuit_2by2}), the state $|x_2x_3x_4\rangle$ becomes $\sum_{j=1}^2\beta_j|2\lambda_j^{-1}\rangle|u_j\rangle$. 

Then we use the $|2\lambda_j^{-1}\rangle$ states in $|x_2x_3\rangle$ as the control register to execute a $Y$ rotation $R_y(\tilde\theta_j)$ on qubit $|x_1\rangle$ with $\tilde\theta_j=2^{1-r}\pi/\lambda_j=2{C}/\lambda_j$, which approximates $\theta_j=2\text{arcsin}({C}/\lambda_j)$. We have previously assumed that $C\le\min_j|\lambda_j|$. Hence we let $r\ge\log_2(2\pi)\approx 2.65$. In general $r$ cannot be too small so as to render the small-angle approximation $\tilde\theta_j$ of $\theta_j$ invalid. At the same time $r$ cannot be too large because the larger $r$ is, the less probable it is to obtain the solution and also finer angles will have to be resolved in the controlled rotation gates, which poses more challenges for implementation. Suppose the minimum angle resolution realizable is $\omega$, then $r\le\log_2(\pi/\omega)$.

\begin{figure}
\includegraphics[scale=0.8]{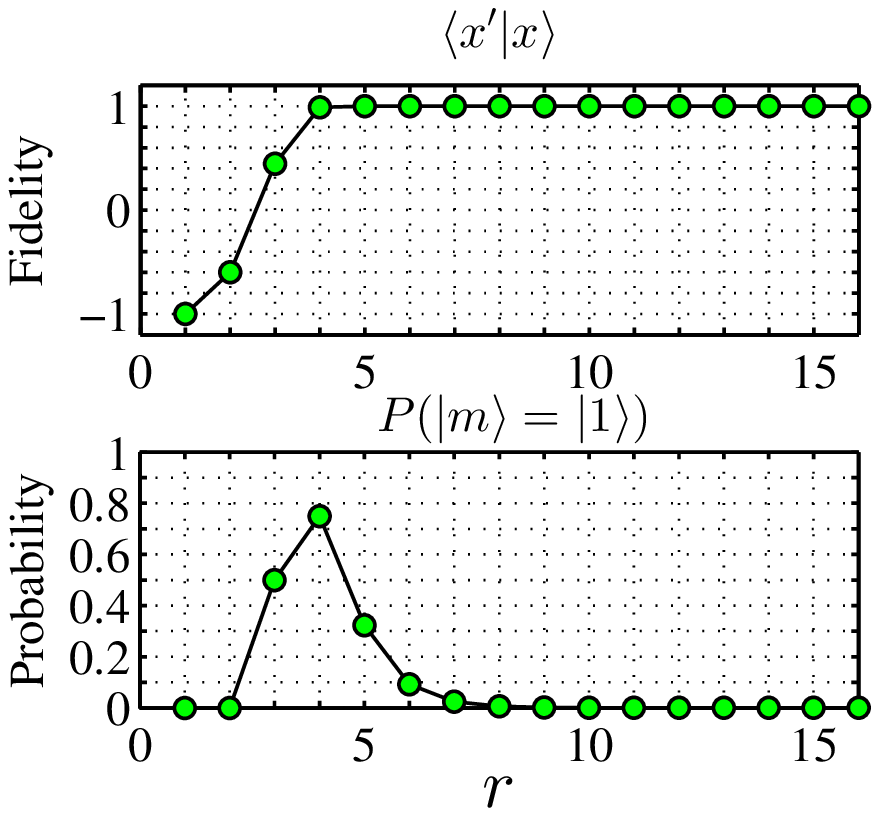}
\caption{Numerical calculation of the fidelity $\langle{x'}|x\rangle$ of the quantum solution and probablity of obtaining $|x'\rangle$ as a function of $r$. $|m\rangle$ is the state $|x_1\rangle$ after measurement (Fig.\ \ref{fig:circuit_2by2}).}
\label{fig:data_2by2}
\end{figure}

Numerical results simulating the circuit with different values of $r$ are shown in Fig.\ \ref{fig:data_2by2}. When the value of $r$ is sufficiently large the fidelity $\langle{x'}|x\rangle$ of the solution converges to 1. Here $|x'\rangle$ is the state of $|x_3x_4\rangle$ when $|x_1\rangle$ is measured to be $|1\rangle$ in the final state. From the previous analysis we see that $|x'\rangle=\cos\tilde\theta_1|u_1\rangle+\sin\tilde\theta_2|u_2\rangle$. $|x\rangle$ corresponds to the analytical solution $\vec{x}=(3/8,-1/8)^T$. The numerical results in Fig.\ \ref{fig:data_2by2} also show that as $r$ grows beyond around a certain point (near $r\approx 4$), the probability of measuring the ancilla bit as $|1\rangle$ decays, which indicates that as $r$ is increased, the solutions obtained in the final state in register $b$ becomes more accurate yet less probable to obtain.

In conclusion, in this work we discuss both general and special case quantum circuit implementation for the algorithm for solving linear systems of equations. Our results may motivate experimentalists with the capability of addressing 4 or more qubits and execute basic quantum gates on their setups to implement the algorithm and verify its results.
$\quad$\\
$\quad$\\
$\quad$\\
\centerline{\it Acknowledgment}
$\quad$\\
This work is dedicated to professor Dudley Herschbach. Also we thank the NSF Center for Quantum Information and Computation for Chemistry, Award number CHE-1037992.

$\quad$\\
\centerline{\it Appendix}
$\quad$\\
The matrix representations of the quantum gates used in this work are the following:
$Y$ and $H$ gates which are the Pauli Y operator and Hadamard gate:

\begin{equation} \label{eq:XYZH}
Y=
\begin{pmatrix}
0 & -i \\
i & 0
\end{pmatrix}, 
H=\frac{1}{\sqrt{2}}
\begin{pmatrix}
1 & 1 \\
1 & -1
\end{pmatrix}.
\end{equation}

\noindent{}The $S$ gate and rotation $R_y$ gate are defined as

\begin{equation} \label{eq:ST}
S=
\begin{pmatrix}
1 & 0 \\
0 & i
\end{pmatrix},
R_y(\theta)=
\begin{pmatrix}
\text{cos}({\theta}/{2}) & -\text{sin}({\theta}/{2}) \\
\text{sin}({\theta}/{2}) & \text{cos}({\theta}/{2})
\end{pmatrix}.
\end{equation}

\bibliographystyle{apsrev4-1}
\bibliography{ref}

\begin{thebibliography}{41}%
\makeatletter
\providecommand \@ifxundefined [1]{%
 \@ifx{#1\undefined}
}%
\providecommand \@ifnum [1]{%
 \ifnum #1\expandafter \@firstoftwo
 \else \expandafter \@secondoftwo
 \fi
}%
\providecommand \@ifx [1]{%
 \ifx #1\expandafter \@firstoftwo
 \else \expandafter \@secondoftwo
 \fi
}%
\providecommand \natexlab [1]{#1}%
\providecommand \enquote  [1]{``#1''}%
\providecommand \bibnamefont  [1]{#1}%
\providecommand \bibfnamefont [1]{#1}%
\providecommand \citenamefont [1]{#1}%
\providecommand \href@noop [0]{\@secondoftwo}%
\providecommand \href [0]{\begingroup \@sanitize@url \@href}%
\providecommand \@href[1]{\@@startlink{#1}\@@href}%
\providecommand \@@href[1]{\endgroup#1\@@endlink}%
\providecommand \@sanitize@url [0]{\catcode `\\12\catcode `\$12\catcode
  `\&12\catcode `\#12\catcode `\^12\catcode `\_12\catcode `\%12\relax}%
\providecommand \@@startlink[1]{}%
\providecommand \@@endlink[0]{}%
\providecommand \url  [0]{\begingroup\@sanitize@url \@url }%
\providecommand \@url [1]{\endgroup\@href {#1}{\urlprefix }}%
\providecommand \urlprefix  [0]{URL }%
\providecommand \Eprint [0]{\href }%
\providecommand \doibase [0]{http://dx.doi.org/}%
\providecommand \selectlanguage [0]{\@gobble}%
\providecommand \bibinfo  [0]{\@secondoftwo}%
\providecommand \bibfield  [0]{\@secondoftwo}%
\providecommand \translation [1]{[#1]}%
\providecommand \BibitemOpen [0]{}%
\providecommand \bibitemStop [0]{}%
\providecommand \bibitemNoStop [0]{.\EOS\space}%
\providecommand \EOS [0]{\spacefactor3000\relax}%
\providecommand \BibitemShut  [1]{\csname bibitem#1\endcsname}%
\let\auto@bib@innerbib\@empty
\bibitem [{\citenamefont {P.W.Shor}(1994)}]{shor94}%
  \BibitemOpen
  \bibfield  {author} {\bibinfo {author} {\bibnamefont {P.W.Shor}},\ }in\
  \href@noop {} {\emph {\bibinfo {booktitle} {Proc. 35th Annu. Symp. Found.
  Comp. Sci.}}},\ \bibinfo {editor} {edited by\ \bibinfo {editor} {\bibnamefont
  {S.Goldwasser}}}\ (\bibinfo  {publisher} {IEEE Computer Society Press},\
  \bibinfo {address} {New York},\ \bibinfo {year} {1994})\ pp.\ \bibinfo
  {pages} {124--134}\BibitemShut {NoStop}%
\bibitem [{\citenamefont {Lloyd}(1996)}]{lloyd96}%
  \BibitemOpen
  \bibfield  {author} {\bibinfo {author} {\bibfnamefont {S.}~\bibnamefont
  {Lloyd}},\ }\href@noop {} {\bibfield  {journal} {\bibinfo  {journal}
  {Science}\ }\textbf {\bibinfo {volume} {273}},\ \bibinfo {pages} {1073}
  (\bibinfo {year} {1996})}\BibitemShut {NoStop}%
\bibitem [{\citenamefont {Abrams}\ and\ \citenamefont
  {Lloyd}(1999)}]{abrams99}%
  \BibitemOpen
  \bibfield  {author} {\bibinfo {author} {\bibfnamefont {D.~S.}\ \bibnamefont
  {Abrams}}\ and\ \bibinfo {author} {\bibfnamefont {S.}~\bibnamefont {Lloyd}},\
  }\href@noop {} {\bibfield  {journal} {\bibinfo  {journal} {Phys. Rev. Lett.}\
  }\textbf {\bibinfo {volume} {83}},\ \bibinfo {pages} {5162} (\bibinfo {year}
  {1999})}\BibitemShut {NoStop}%
\bibitem [{\citenamefont {Abrams}\ and\ \citenamefont
  {Lloyd}(1997)}]{abrams97}%
  \BibitemOpen
  \bibfield  {author} {\bibinfo {author} {\bibfnamefont {D.~S.}\ \bibnamefont
  {Abrams}}\ and\ \bibinfo {author} {\bibfnamefont {S.}~\bibnamefont {Lloyd}},\
  }\href@noop {} {\bibfield  {journal} {\bibinfo  {journal} {Phys. Rev. Lett.}\
  }\textbf {\bibinfo {volume} {79}},\ \bibinfo {pages} {2586} (\bibinfo {year}
  {1997})}\BibitemShut {NoStop}%
\bibitem [{\citenamefont {Papageorgiou}\ \emph {et~al.}(94v2)\citenamefont
  {Papageorgiou}, \citenamefont {Petras}, \citenamefont {Traub},\ and\
  \citenamefont {Zhang}}]{traub10}%
  \BibitemOpen
  \bibfield  {author} {\bibinfo {author} {\bibfnamefont {A.}~\bibnamefont
  {Papageorgiou}}, \bibinfo {author} {\bibfnamefont {I.}~\bibnamefont
  {Petras}}, \bibinfo {author} {\bibfnamefont {J.~F.}\ \bibnamefont {Traub}}, \
  and\ \bibinfo {author} {\bibfnamefont {C.}~\bibnamefont {Zhang}},\
  }\href@noop {} {\  (\bibinfo {year} {arXiv:1008.4294v2})}\BibitemShut
  {NoStop}%
\bibitem [{\citenamefont {Papageorgiou}\ and\ \citenamefont
  {Zhang}(2012)}]{zhang10}%
  \BibitemOpen
  \bibfield  {author} {\bibinfo {author} {\bibfnamefont {A.}~\bibnamefont
  {Papageorgiou}}\ and\ \bibinfo {author} {\bibfnamefont {C.}~\bibnamefont
  {Zhang}},\ }\href@noop {} {\bibfield  {journal} {\bibinfo  {journal} {Quantum
  Information Processing}\ }\textbf {\bibinfo {volume} {11}},\ \bibinfo {pages}
  {541} (\bibinfo {year} {2012})}\BibitemShut {NoStop}%
\bibitem [{\citenamefont {Wang}\ \emph {et~al.}(2008)\citenamefont {Wang},
  \citenamefont {Kais}, \citenamefont {Aspuru-Guzik},\ and\ \citenamefont
  {Hoffmann}}]{kais08}%
  \BibitemOpen
  \bibfield  {author} {\bibinfo {author} {\bibfnamefont {H.}~\bibnamefont
  {Wang}}, \bibinfo {author} {\bibfnamefont {S.}~\bibnamefont {Kais}}, \bibinfo
  {author} {\bibfnamefont {A.}~\bibnamefont {Aspuru-Guzik}}, \ and\ \bibinfo
  {author} {\bibfnamefont {M.~R.}\ \bibnamefont {Hoffmann}},\ }\href@noop {}
  {\bibfield  {journal} {\bibinfo  {journal} {Phys. Chem. Chem. Phys.}\
  }\textbf {\bibinfo {volume} {10}},\ \bibinfo {pages} {5388} (\bibinfo {year}
  {2008})}\BibitemShut {NoStop}%
\bibitem [{\citenamefont {Aspuru-Guzik}\ \emph {et~al.}(2005)\citenamefont
  {Aspuru-Guzik}, \citenamefont {Dutoi}, \citenamefont {Love},\ and\
  \citenamefont {Head-Gordon}}]{alan05}%
  \BibitemOpen
  \bibfield  {author} {\bibinfo {author} {\bibfnamefont {A.}~\bibnamefont
  {Aspuru-Guzik}}, \bibinfo {author} {\bibfnamefont {A.~D.}\ \bibnamefont
  {Dutoi}}, \bibinfo {author} {\bibfnamefont {P.~J.}\ \bibnamefont {Love}}, \
  and\ \bibinfo {author} {\bibfnamefont {M.}~\bibnamefont {Head-Gordon}},\
  }\href@noop {} {\bibfield  {journal} {\bibinfo  {journal} {Science}\ }\textbf
  {\bibinfo {volume} {379}},\ \bibinfo {pages} {1704} (\bibinfo {year}
  {2005})}\BibitemShut {NoStop}%
\bibitem [{\citenamefont {Lidar}\ and\ \citenamefont {Wang}(1999)}]{lidar99}%
  \BibitemOpen
  \bibfield  {author} {\bibinfo {author} {\bibfnamefont {D.}~\bibnamefont
  {Lidar}}\ and\ \bibinfo {author} {\bibfnamefont {H.}~\bibnamefont {Wang}},\
  }\href@noop {} {\bibfield  {journal} {\bibinfo  {journal} {Phys. Rev. E}\
  }\textbf {\bibinfo {volume} {59}},\ \bibinfo {pages} {2429} (\bibinfo {year}
  {1999})}\BibitemShut {NoStop}%
\bibitem [{\citenamefont {Wang}\ \emph {et~al.}()\citenamefont {Wang},
  \citenamefont {Ashhab},\ and\ \citenamefont {Nori}}]{nori11a}%
  \BibitemOpen
  \bibfield  {author} {\bibinfo {author} {\bibfnamefont {H.}~\bibnamefont
  {Wang}}, \bibinfo {author} {\bibfnamefont {S.}~\bibnamefont {Ashhab}}, \ and\
  \bibinfo {author} {\bibfnamefont {F.}~\bibnamefont {Nori}},\ }\href@noop {}
  {\bibinfo  {journal} {Phys. Rev. E}\ ,\ \bibinfo {pages}
  {arXiv:1108.5902}}\BibitemShut {NoStop}%
\bibitem [{\citenamefont {You}\ and\ \citenamefont {Nori}(2011)}]{nori11b}%
  \BibitemOpen
\bibfield  {journal} {  }\bibfield  {author} {\bibinfo {author} {\bibfnamefont
  {J.}~\bibnamefont {You}}\ and\ \bibinfo {author} {\bibfnamefont
  {F.}~\bibnamefont {Nori}},\ }\href@noop {} {\bibfield  {journal} {\bibinfo
  {journal} {Nature}\ }\textbf {\bibinfo {volume} {474}},\ \bibinfo {pages}
  {589} (\bibinfo {year} {2011})}\BibitemShut {NoStop}%
\bibitem [{\citenamefont {You}\ and\ \citenamefont {Nori}(2005)}]{nori05}%
  \BibitemOpen
  \bibfield  {author} {\bibinfo {author} {\bibfnamefont {J.}~\bibnamefont
  {You}}\ and\ \bibinfo {author} {\bibfnamefont {F.}~\bibnamefont {Nori}},\
  }\href@noop {} {\bibfield  {journal} {\bibinfo  {journal} {Phys. Today}\
  }\textbf {\bibinfo {volume} {58}},\ \bibinfo {pages} {42} (\bibinfo {year}
  {2005})}\BibitemShut {NoStop}%
\bibitem [{\citenamefont {Buluta}\ and\ \citenamefont {Nori}(2009)}]{nori09}%
  \BibitemOpen
  \bibfield  {author} {\bibinfo {author} {\bibfnamefont {I.}~\bibnamefont
  {Buluta}}\ and\ \bibinfo {author} {\bibfnamefont {F.}~\bibnamefont {Nori}},\
  }\href@noop {} {\bibfield  {journal} {\bibinfo  {journal} {Science}\ }\textbf
  {\bibinfo {volume} {326}} (\bibinfo {year} {2009})}\BibitemShut {NoStop}%
\bibitem [{\citenamefont {Dowling}(2006)}]{dowling06}%
  \BibitemOpen
  \bibfield  {author} {\bibinfo {author} {\bibfnamefont {J.}~\bibnamefont
  {Dowling}},\ }\href@noop {} {\bibfield  {journal} {\bibinfo  {journal}
  {Nature}\ }\textbf {\bibinfo {volume} {439}},\ \bibinfo {pages} {919}
  (\bibinfo {year} {2006})}\BibitemShut {NoStop}%
\bibitem [{\citenamefont {Veis}\ and\ \citenamefont {Pittner}(2010)}]{veis10}%
  \BibitemOpen
  \bibfield  {author} {\bibinfo {author} {\bibfnamefont {L.}~\bibnamefont
  {Veis}}\ and\ \bibinfo {author} {\bibfnamefont {J.}~\bibnamefont {Pittner}},\
  }\href@noop {} {\bibfield  {journal} {\bibinfo  {journal} {J. Chem. Phys.}\
  }\textbf {\bibinfo {volume} {133}} (\bibinfo {year} {2010})}\BibitemShut
  {NoStop}%
\bibitem [{\citenamefont {Veis}\ \emph {et~al.}(2011)\citenamefont {Veis},
  \citenamefont {Visnak}, \citenamefont {Fleig}, \citenamefont {Knecht},
  \citenamefont {Saue}, \citenamefont {Visscher},\ and\ \citenamefont
  {Pittner}}]{veis11}%
  \BibitemOpen
  \bibfield  {author} {\bibinfo {author} {\bibfnamefont {L.}~\bibnamefont
  {Veis}}, \bibinfo {author} {\bibfnamefont {J.}~\bibnamefont {Visnak}},
  \bibinfo {author} {\bibfnamefont {T.}~\bibnamefont {Fleig}}, \bibinfo
  {author} {\bibfnamefont {S.}~\bibnamefont {Knecht}}, \bibinfo {author}
  {\bibfnamefont {T.}~\bibnamefont {Saue}}, \bibinfo {author} {\bibfnamefont
  {L.}~\bibnamefont {Visscher}}, \ and\ \bibinfo {author} {\bibfnamefont
  {J.}~\bibnamefont {Pittner}},\ }\href@noop {} {\bibfield  {journal} {\bibinfo
   {journal} {arXiv:1111.3490v1}\ } (\bibinfo {year} {2011})}\BibitemShut
  {NoStop}%
\bibitem [{\citenamefont {Harrow}\ \emph {et~al.}(2009)\citenamefont {Harrow},
  \citenamefont {Hassidim},\ and\ \citenamefont {Lloyd}}]{lloyd09}%
  \BibitemOpen
  \bibfield  {author} {\bibinfo {author} {\bibfnamefont {A.~W.}\ \bibnamefont
  {Harrow}}, \bibinfo {author} {\bibfnamefont {A.}~\bibnamefont {Hassidim}}, \
  and\ \bibinfo {author} {\bibfnamefont {S.}~\bibnamefont {Lloyd}},\
  }\href@noop {} {\bibfield  {journal} {\bibinfo  {journal} {Phys. Rev. Lett.}\
  }\textbf {\bibinfo {volume} {15}},\ \bibinfo {pages} {150502} (\bibinfo
  {year} {2009})}\BibitemShut {NoStop}%
\bibitem [{\citenamefont {Nielsen}\ and\ \citenamefont
  {Chuang}(2000)}]{nielsen00}%
  \BibitemOpen
  \bibfield  {author} {\bibinfo {author} {\bibfnamefont {M.~A.}\ \bibnamefont
  {Nielsen}}\ and\ \bibinfo {author} {\bibfnamefont {I.~L.}\ \bibnamefont
  {Chuang}},\ }\href@noop {} {\emph {\bibinfo {title} {Quantum Computation and
  Quantum Information}}}\ (\bibinfo  {publisher} {Cambridge University Press},\
  \bibinfo {address} {Cambridge, United Kingdom},\ \bibinfo {year}
  {2000})\BibitemShut {NoStop}%
\bibitem [{\citenamefont {Shewchuk}(1994)}]{shewchuk94}%
  \BibitemOpen
  \bibfield  {author} {\bibinfo {author} {\bibfnamefont {J.~R.}\ \bibnamefont
  {Shewchuk}},\ }\href@noop {} {\emph {\bibinfo {title} {An introduction to the
  conjugate gradient method without the agonizing pain}}},\ \bibinfo {type}
  {Tech. Rep.}\ \bibinfo {number} {CMU-CS-94-125}\ (\bibinfo  {institution}
  {School of Computer Science, Carnegie Mellon University},\ \bibinfo {address}
  {Pittsburgh, Pennsylvania},\ \bibinfo {year} {1994})\BibitemShut {NoStop}%
\bibitem [{\citenamefont {Childs}(2009)}]{childs09}%
  \BibitemOpen
  \bibfield  {author} {\bibinfo {author} {\bibfnamefont {A.~M.}\ \bibnamefont
  {Childs}},\ }\href@noop {} {\bibfield  {journal} {\bibinfo  {journal} {Nature
  Phys.}\ }\textbf {\bibinfo {volume} {5}},\ \bibinfo {pages} {861} (\bibinfo
  {year} {2009})}\BibitemShut {NoStop}%
\bibitem [{\citenamefont {M\"{o}tt\"{o}nen}\ \emph {et~al.}(2004)\citenamefont
  {M\"{o}tt\"{o}nen}, \citenamefont {Vartiainen}, \citenamefont {Bergholm},\
  and\ \citenamefont {Salomaa}}]{MVBS04}%
  \BibitemOpen
  \bibfield  {author} {\bibinfo {author} {\bibfnamefont {M.}~\bibnamefont
  {M\"{o}tt\"{o}nen}}, \bibinfo {author} {\bibfnamefont {J.~J.}\ \bibnamefont
  {Vartiainen}}, \bibinfo {author} {\bibfnamefont {V.}~\bibnamefont
  {Bergholm}}, \ and\ \bibinfo {author} {\bibfnamefont {M.~M.}\ \bibnamefont
  {Salomaa}},\ }\href@noop {} {\bibfield  {journal} {\bibinfo  {journal} {Phys.
  Rev. Lett.}\ }\textbf {\bibinfo {volume} {93}} (\bibinfo {year}
  {2004})}\BibitemShut {NoStop}%
\bibitem [{\citenamefont {Bergholm}\ \emph {et~al.}(2005)\citenamefont
  {Bergholm}, \citenamefont {Vartiainen}, \citenamefont {M\"{o}tt\"{o}nen},\
  and\ \citenamefont {Salomaa}}]{BVMS05}%
  \BibitemOpen
  \bibfield  {author} {\bibinfo {author} {\bibfnamefont {V.}~\bibnamefont
  {Bergholm}}, \bibinfo {author} {\bibfnamefont {J.~J.}\ \bibnamefont
  {Vartiainen}}, \bibinfo {author} {\bibfnamefont {M.}~\bibnamefont
  {M\"{o}tt\"{o}nen}}, \ and\ \bibinfo {author} {\bibfnamefont {M.~M.}\
  \bibnamefont {Salomaa}},\ }\href@noop {} {\bibfield  {journal} {\bibinfo
  {journal} {Phys. Rev. Lett. A}\ }\textbf {\bibinfo {volume} {71}} (\bibinfo
  {year} {2005})}\BibitemShut {NoStop}%
\bibitem [{\citenamefont {Plesch}\ and\ \citenamefont {\v{C}aslav
  Brukner}(2011)}]{PB11}%
  \BibitemOpen
  \bibfield  {author} {\bibinfo {author} {\bibfnamefont {M.}~\bibnamefont
  {Plesch}}\ and\ \bibinfo {author} {\bibnamefont {\v{C}aslav Brukner}},\
  }\href@noop {} {\  (\bibinfo {year} {2011})},\ \bibinfo {note}
  {arXiv:1003.5760v2 [quant-ph]}\BibitemShut {NoStop}%
\bibitem [{\citenamefont {Zalka}(1998)}]{Z98}%
  \BibitemOpen
  \bibfield  {author} {\bibinfo {author} {\bibfnamefont {C.}~\bibnamefont
  {Zalka}},\ }\href@noop {} {\bibfield  {journal} {\bibinfo  {journal} {Proc.
  R. Soc. London Ser. A}\ }\textbf {\bibinfo {volume} {454}} (\bibinfo {year}
  {1998})}\BibitemShut {NoStop}%
\bibitem [{\citenamefont {Grover}\ and\ \citenamefont {Rudolph}(2002)}]{GR02}%
  \BibitemOpen
  \bibfield  {author} {\bibinfo {author} {\bibfnamefont {L.}~\bibnamefont
  {Grover}}\ and\ \bibinfo {author} {\bibfnamefont {T.}~\bibnamefont
  {Rudolph}},\ }\href@noop {} {\  (\bibinfo {year} {2002})},\ \bibinfo {note}
  {arXiv:quant-ph/0208112v1}\BibitemShut {NoStop}%
\bibitem [{\citenamefont {Kaye}\ and\ \citenamefont {Mosca}(2004)}]{KM04}%
  \BibitemOpen
  \bibfield  {author} {\bibinfo {author} {\bibfnamefont {P.}~\bibnamefont
  {Kaye}}\ and\ \bibinfo {author} {\bibfnamefont {M.}~\bibnamefont {Mosca}},\
  }\href@noop {} {\  (\bibinfo {year} {2004})},\ \bibinfo {note}
  {arXiv:quant-ph/0407102v1}\BibitemShut {NoStop}%
\bibitem [{\citenamefont {Ward}\ \emph {et~al.}(2009)\citenamefont {Ward},
  \citenamefont {Kassal},\ and\ \citenamefont {Aspuru-Guzik}}]{WKA09}%
  \BibitemOpen
  \bibfield  {author} {\bibinfo {author} {\bibfnamefont {N.~J.}\ \bibnamefont
  {Ward}}, \bibinfo {author} {\bibfnamefont {I.}~\bibnamefont {Kassal}}, \ and\
  \bibinfo {author} {\bibfnamefont {A.}~\bibnamefont {Aspuru-Guzik}},\
  }\href@noop {} {\bibfield  {journal} {\bibinfo  {journal} {J. Chem. Phys.}\
  }\textbf {\bibinfo {volume} {130}} (\bibinfo {year} {2009})}\BibitemShut
  {NoStop}%
\bibitem [{\citenamefont {Aharonov}\ and\ \citenamefont
  {Ta-Shma}(2003)}]{AT03}%
  \BibitemOpen
  \bibfield  {author} {\bibinfo {author} {\bibfnamefont {D.}~\bibnamefont
  {Aharonov}}\ and\ \bibinfo {author} {\bibfnamefont {A.}~\bibnamefont
  {Ta-Shma}},\ }\href@noop {} {\bibfield  {journal} {\bibinfo  {journal}
  {Proceedings of the 35th ACM symposium on theory of computing}\ } (\bibinfo
  {year} {2003})},\ \bibinfo {note} {arXiv:quant-ph/0301023}\BibitemShut
  {NoStop}%
\bibitem [{\citenamefont {Aharonov}\ and\ \citenamefont
  {Ta-Shma}(2007)}]{AT07}%
  \BibitemOpen
  \bibfield  {author} {\bibinfo {author} {\bibfnamefont {D.}~\bibnamefont
  {Aharonov}}\ and\ \bibinfo {author} {\bibfnamefont {A.}~\bibnamefont
  {Ta-Shma}},\ }\href@noop {} {\bibfield  {journal} {\bibinfo  {journal} {SIAM
  J. Comput.}\ }\textbf {\bibinfo {volume} {37}} (\bibinfo {year}
  {2007})}\BibitemShut {NoStop}%
\bibitem [{\citenamefont {Childs}\ and\ \citenamefont
  {Kothari}(2009)}]{childs10}%
  \BibitemOpen
  \bibfield  {author} {\bibinfo {author} {\bibfnamefont {A.~M.}\ \bibnamefont
  {Childs}}\ and\ \bibinfo {author} {\bibfnamefont {R.}~\bibnamefont
  {Kothari}},\ }\href@noop {} {\bibfield  {journal} {\bibinfo  {journal}
  {Quantum Information and Computation}\ }\textbf {\bibinfo {volume} {10}},\
  \bibinfo {pages} {869} (\bibinfo {year} {2009})}\BibitemShut {NoStop}%
\bibitem [{\citenamefont {Childs}\ \emph {et~al.}(2003)\citenamefont {Childs},
  \citenamefont {Cleve}, \citenamefont {Deotto}, \citenamefont {Farhi},
  \citenamefont {Gutmann},\ and\ \citenamefont {Spielman}}]{childs03}%
  \BibitemOpen
  \bibfield  {author} {\bibinfo {author} {\bibfnamefont {A.~M.}\ \bibnamefont
  {Childs}}, \bibinfo {author} {\bibfnamefont {R.}~\bibnamefont {Cleve}},
  \bibinfo {author} {\bibfnamefont {E.}~\bibnamefont {Deotto}}, \bibinfo
  {author} {\bibfnamefont {E.}~\bibnamefont {Farhi}}, \bibinfo {author}
  {\bibfnamefont {S.}~\bibnamefont {Gutmann}}, \ and\ \bibinfo {author}
  {\bibfnamefont {D.~A.}\ \bibnamefont {Spielman}},\ }in\ \href@noop {} {\emph
  {\bibinfo {booktitle} {Proceedings of the 35th ACM Symposium on Theory of
  Computing}}}\ (\bibinfo {year} {2003})\ pp.\ \bibinfo {pages} {59--68},\
  \bibinfo {note} {arXiv:quant-ph/0209131}\BibitemShut {NoStop}%
\bibitem [{\citenamefont {Berry}\ \emph {et~al.}(2007)\citenamefont {Berry},
  \citenamefont {Ahokas}, \citenamefont {Cleve},\ and\ \citenamefont
  {Sanders}}]{berry07}%
  \BibitemOpen
  \bibfield  {author} {\bibinfo {author} {\bibfnamefont {D.~W.}\ \bibnamefont
  {Berry}}, \bibinfo {author} {\bibfnamefont {G.}~\bibnamefont {Ahokas}},
  \bibinfo {author} {\bibfnamefont {R.}~\bibnamefont {Cleve}}, \ and\ \bibinfo
  {author} {\bibfnamefont {B.~C.}\ \bibnamefont {Sanders}},\ }\href@noop {}
  {\bibfield  {journal} {\bibinfo  {journal} {Communications in Mathematical
  Physics}\ }\textbf {\bibinfo {volume} {270}},\ \bibinfo {pages} {359}
  (\bibinfo {year} {2007})}\BibitemShut {NoStop}%
\bibitem [{\citenamefont {Wiebe}\ \emph {et~al.}(2011)\citenamefont {Wiebe},
  \citenamefont {Berry}, \citenamefont {H{\o}yer},\ and\ \citenamefont
  {Sanders}}]{wiebe11}%
  \BibitemOpen
  \bibfield  {author} {\bibinfo {author} {\bibfnamefont {N.}~\bibnamefont
  {Wiebe}}, \bibinfo {author} {\bibfnamefont {D.~W.}\ \bibnamefont {Berry}},
  \bibinfo {author} {\bibfnamefont {P.}~\bibnamefont {H{\o}yer}}, \ and\
  \bibinfo {author} {\bibfnamefont {B.~C.}\ \bibnamefont {Sanders}},\
  }\href@noop {} {\bibfield  {journal} {\bibinfo  {journal} {Journal of Physics
  A}\ }\textbf {\bibinfo {volume} {44}} (\bibinfo {year} {2011})},\ \bibinfo
  {note} {arXiv:1011.3489}\BibitemShut {NoStop}%
\bibitem [{\citenamefont {Childs}\ and\ \citenamefont {Wiebe}(2012)}]{wiebe12}%
  \BibitemOpen
  \bibfield  {author} {\bibinfo {author} {\bibfnamefont {A.~M.}\ \bibnamefont
  {Childs}}\ and\ \bibinfo {author} {\bibfnamefont {N.}~\bibnamefont {Wiebe}},\
  }\href@noop {} {\bibfield  {journal} {\bibinfo  {journal} {Quantum
  information and computation}\ }\textbf {\bibinfo {volume} {12}},\ \bibinfo
  {pages} {901} (\bibinfo {year} {2012})}\BibitemShut {NoStop}%
\bibitem [{\citenamefont {Childs}(2010)}]{childs09b}%
  \BibitemOpen
  \bibfield  {author} {\bibinfo {author} {\bibfnamefont {A.~M.}\ \bibnamefont
  {Childs}},\ }\href@noop {} {\bibfield  {journal} {\bibinfo  {journal} {Comm.
  Math. Phys.}\ }\textbf {\bibinfo {volume} {294}},\ \bibinfo {pages} {581}
  (\bibinfo {year} {2010})},\ \bibinfo {note} {arXiv:0810.0312
  [quant-ph]}\BibitemShut {NoStop}%
\bibitem [{\citenamefont {Berry}\ and\ \citenamefont {Childs}(2012)}]{berry12}%
  \BibitemOpen
  \bibfield  {author} {\bibinfo {author} {\bibfnamefont {D.~W.}\ \bibnamefont
  {Berry}}\ and\ \bibinfo {author} {\bibfnamefont {A.~M.}\ \bibnamefont
  {Childs}},\ }\href@noop {} {\bibfield  {journal} {\bibinfo  {journal}
  {Quantum information and computation}\ }\textbf {\bibinfo {volume} {12}}
  (\bibinfo {year} {2012})},\ \bibinfo {note} {arXiv:0910.4157
  [quant-ph]}\BibitemShut {NoStop}%
\bibitem [{\citenamefont {Cao}\ \emph {et~al.}(2013)\citenamefont {Cao},
  \citenamefont {Papageorgiou}, \citenamefont {Petras}, \citenamefont {Traub},\
  and\ \citenamefont {Kais}}]{cao13}%
  \BibitemOpen
  \bibfield  {author} {\bibinfo {author} {\bibfnamefont {Y.}~\bibnamefont
  {Cao}}, \bibinfo {author} {\bibfnamefont {A.}~\bibnamefont {Papageorgiou}},
  \bibinfo {author} {\bibfnamefont {I.}~\bibnamefont {Petras}}, \bibinfo
  {author} {\bibfnamefont {J.}~\bibnamefont {Traub}}, \ and\ \bibinfo {author}
  {\bibfnamefont {S.}~\bibnamefont {Kais}},\ }\href@noop {} {\bibfield
  {journal} {\bibinfo  {journal} {New Journal of Physics}\ }\textbf {\bibinfo
  {volume} {15}} (\bibinfo {year} {2013})}\BibitemShut {NoStop}%
\bibitem [{\citenamefont {Cleve}\ \emph {et~al.}(2009)\citenamefont {Cleve},
  \citenamefont {Gottesman}, \citenamefont {Mosca}, \citenamefont {Somma},\
  and\ \citenamefont {Yonge-Mallo}}]{cleve09}%
  \BibitemOpen
  \bibfield  {author} {\bibinfo {author} {\bibfnamefont {R.}~\bibnamefont
  {Cleve}}, \bibinfo {author} {\bibfnamefont {D.}~\bibnamefont {Gottesman}},
  \bibinfo {author} {\bibfnamefont {M.}~\bibnamefont {Mosca}}, \bibinfo
  {author} {\bibfnamefont {R.}~\bibnamefont {Somma}}, \ and\ \bibinfo {author}
  {\bibfnamefont {D.}~\bibnamefont {Yonge-Mallo}},\ }in\ \href@noop {} {\emph
  {\bibinfo {booktitle} {Proc. 41st Ann. Symp. on Theory of Computing}}}\
  (\bibinfo {year} {2009})\ pp.\ \bibinfo {pages} {409--416}\BibitemShut
  {NoStop}%
\bibitem [{\citenamefont {Berry}\ \emph {et~al.}(2013)\citenamefont {Berry},
  \citenamefont {Cleve},\ and\ \citenamefont {Somma}}]{berry13}%
  \BibitemOpen
  \bibfield  {author} {\bibinfo {author} {\bibfnamefont {D.}~\bibnamefont
  {Berry}}, \bibinfo {author} {\bibfnamefont {R.}~\bibnamefont {Cleve}}, \ and\
  \bibinfo {author} {\bibfnamefont {R.}~\bibnamefont {Somma}},\ }\href@noop {}
  {\  (\bibinfo {year} {2013})},\ \bibinfo {note}
  {www.dominicberry.org/presentations/polylog.pptx‎}\BibitemShut {NoStop}%
\bibitem [{\citenamefont {Daskin}\ and\ \citenamefont
  {Kais}(2011{\natexlab{a}})}]{daskin11a}%
  \BibitemOpen
  \bibfield  {author} {\bibinfo {author} {\bibfnamefont {A.}~\bibnamefont
  {Daskin}}\ and\ \bibinfo {author} {\bibfnamefont {S.}~\bibnamefont {Kais}},\
  }\href@noop {} {\bibfield  {journal} {\bibinfo  {journal} {J. Chem. Phys.}\
  }\textbf {\bibinfo {volume} {134}} (\bibinfo {year}
  {2011}{\natexlab{a}})}\BibitemShut {NoStop}%
\bibitem [{\citenamefont {Daskin}\ and\ \citenamefont
  {Kais}(2011{\natexlab{b}})}]{daskin11b}%
  \BibitemOpen
  \bibfield  {author} {\bibinfo {author} {\bibfnamefont {A.}~\bibnamefont
  {Daskin}}\ and\ \bibinfo {author} {\bibfnamefont {S.}~\bibnamefont {Kais}},\
  }\href@noop {} {\bibfield  {journal} {\bibinfo  {journal} {Mol. Phys.}\
  }\textbf {\bibinfo {volume} {109}},\ \bibinfo {pages} {761} (\bibinfo {year}
  {2011}{\natexlab{b}})}\BibitemShut {NoStop}%
\end{thebibliography}%

\end{document}